\documentclass[twocolumn,showpacs,amsmath,amssymb,superscriptaddress]{revtex4-1}
\usepackage{graphicx}
\usepackage{dcolumn}
\usepackage{bm}
\usepackage{hyperref}
\usepackage{epsf}
\usepackage{float} 
\usepackage{soul}
\newcommand{\eq}[1]{(\ref{#1})}
\newcommand{\la}{\label}
\newcommand{\bea}{\begin{eqnarray}}
\newcommand{\eea}{\end{eqnarray}}
\newcommand{\beq}{\begin{equation}}
\newcommand{\eeq}{\end{equation}}
\newcommand{\be}{\begin{equation}}
\newcommand{\ee}{\end{equation}}
\newcommand{\ii}{{\rm{i}}}

\newcommand{\pp}{{\rm p}}
\newcommand{\p}{\partial}

\newcommand{\erfc}{{\rm erfc}}

\def\XXint#1#2#3{{\setbox0=\hbox{$#1{#2#3}{\int}$ }
\vcenter{\hbox{$#2#3$ }}\kern-.5\wd0}}

\usepackage{amsmath}
\usepackage{amsfonts}
\usepackage{varioref}
\usepackage{xcolor}
\begin{document}

\title{Singular Behavior At The Edge of Laughlin States}

\author{T. Can}
\affiliation{Department of Physics, University of Chicago, 929 57th St, Chicago, IL 60637}

\author{P. J. Forrester}
\affiliation{Department of Mathematics and Statistics, The University of Melbourne, Victoria 3010, Australia}

\author{G. T\'ellez}
\affiliation{Departamento de F\'isica, Universidad de Los Andes, Bogot\'a, Colombia} 

\author{P. Wiegmann}
\affiliation{Department of Physics, University of Chicago, 929 57th St, Chicago, IL 60637}

\date{\today}

%

\date{\today}

\begin{abstract}
A distinguishing feature of fractional quantum Hall (FQH) states is a singular behavior of equilibrium densities at boundaries. In contrast to states at integer filling fraction, such quantum liquids posses an additional  dipole moment localized near edges. It enters observable quantities such as universal  dispersion of edge states and Lorentz shear stress. For a Laughlin state, this behavior is seen as a peak, or overshoot, in the single particle density near the edge, reflecting a general tendency of electrons in FQH states to cluster near edges.   We compute the singular edge behavior  of the one particle
density by a perturbative expansion carried out around a completely
filled Landau level. This correction is shown to  fully capture the dipole moment  and the major features of the overshoot  observed numerically. Furthermore, it exhibits the Stokes phenomenon with the Stokes line at the boundary of the droplet,  decaying like a Gaussian inside and outside the liquid with different decay lengths. In the limit of vanishing magnetic length the shape of the overshoot is a singular double layer with a capacity that is a universal function of the filling fraction. Finally, we derive the edge dipole moment of Pfaffian FQH states. The result suggests an explicit connection between the magnitude of the dipole moment and the bulk odd viscosity.
\end{abstract}


\maketitle

\section{Introduction}
\la{I}
One of the distinct features of fractional quantum Hall (FQH) states is the non-monotonic behavior of static correlation functions at short distances. Such behavior has physical consequences. One of the most familiar is the magneto-roton minimum in the
dispersion curve of the gapped collective excitations. This was first demonstrated for Laughlin states (states with the filling fraction equal to the inverse of an odd integer \cite{Laughlin}) using a variational formula for the energy spectrum which  involves the static structure factor of the FQH ground state \cite{GMP}. The dip in the dispersion curve was attributed to a peak which occurs in the numerically computed static structure factor. Numerical studies show that this peak is followed by damped oscillatory features \cite{Caillol}, already apparent in the $\nu=1/3$ state \cite{GMP}. That the magneto-roton minimum disappears for the integer quantum Hall (IQH) state can also be deduced directly from its static structure factor, which, in this case ($\nu=1$) is a monotonically increasing function of momentum.
The pair correlation
function in coordinate
space possesses the same structure as the static structure factor (the
Fourier transform of the pair density correlation function). The pair correlation
function oscillates as separation between points decreases,
then rises up before vanishing at zero separation \cite{GMP}.

\begin{figure}[h]
\includegraphics[scale=.6]{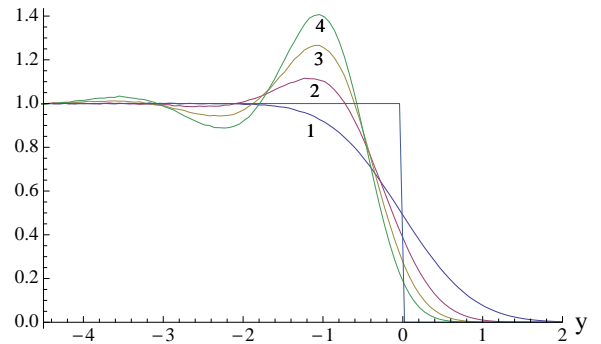}
\caption{Density profile near the edge of the droplet  \(\rho_\beta(y)/\bar\rho\),
labelled by the value of \(\beta\) (equal to the inverse filling fraction \(\nu^{-1}\) for odd integer values). For reference, the support of the droplet
is plotted as a step function. Data obtained by numerical simulation \cite{Shytov}. Distance \(y\) is measured in units of \(\ell\).}
\label{Fig1}
\end{figure}

Similar non-monotonic behavior is seen in the  one particle density of FQH states, which displays a prominent peak, or ``overshoot", at the edge \cite{unrecon}. This behavior has been observed numerically in the  Laughlin states   \cite{Halperin, Morf, Ciftja, Tellez}. This is in stark contrast to the  particle density of the IQH state, which monotonically decreases with distance from the center of mass, as seen in Fig.~\ref{Fig1}. Consequently, states with a fractional filling factor possess a dipole moment, as well as higher order moments,  additional to the integer case and localized near the edge.

The overshoot in the one particle density is seen to have noticeable physical consequences. Recently, it has been  shown that the double layer is an essential ingredient in the the theory of edge waves that supports fractionally charged edge solitons \cite{PW}. In the same paper, it was  conjectured that in the limit of vanishing magnetic
length the edge dipole moment is in fact a  \emph{double layer} with a  capacity
which is a universal function of the filling fraction \cite{PW}. The universal properties of the edge dipole are also closely linked to the Lorentz shear stress, an intrinsic property of the bulk \cite{PW13, FDMH09}. 

There is a very good reason to think that these two peaks, or overshoots, are closely related. Indeed, the pair correlation function in coordinate space is equivalent to the one particle density around a hole, or puncture, which is effectively an edge with large curvature. Thus both features can be understood as a tendency of electrons in the FQH state to cluster at inner edges of the  droplet.

The overshoot, in the density and pair correlation function, should be understood as a key feature of the FQH state, one that is conspicuously absent in the IQH state. 

While the significance of such edge singularities is now appreciated, an understanding of them has thus far relied almost entirely on numerics. Of the handful of analytic works in the literature, we refer to \cite{Girvin,Jancovici} for attempts to capture this behavior in the bulk pair correlation function and \cite{Zabrodin, Forrester} in relation to the boundary overshoot. 

The goal of this article is to shed light on the density overshoot at the edge. We are primarily concerned with the analytical structure of the edge density. We find that  the most reliable approach is to  treat the inverse filling fraction \(\nu^{-1} = \beta\) as a parameter in Laughlin's wave function and to expand the edge density about its known value for the completely filled Landau level $\beta = 1$ to first order in the perturbation variable $\beta - 1$. This approach has been successfully executed  by Jancovici in Ref. \cite{Jancovici} for the bulk pair correlation  function. The leading correction is shown to herald the overshoot for the Laughlin states. Since the  2D one component plasma  is not believed to have a phase transition in the domain of \(\beta\) relevant for FQH states, we assume the results near \(\beta = 1\) can be adiabatically connected to the \(\beta = 3, 5...\), and specifically that the \(\beta - 1\) term evolves smoothly into the overshoot seen in the Laughlin state. This argument is supported by numerical data.

The overshoot at the boundary and subsequent oscillatory features (not captured by the first order correction) extended toward the bulk of the droplet could be assigned to a tendency toward crystallization as $\beta$ increases. The ``crystallization" is more pronounced in the vicinity of the boundary.  

In this paper, we focus on Laughlin states, and briefly discuss possible extensions to other FQH states. We expect more complicated FQH states also feature singular behavior on edges, and as a first step toward understanding their structure we compute the edge dipole moment for Pfaffian states. In Sec. \ref{sec2}, we present our main results, the derivation of which is summarized in the subsequent sections. Further detailed derivations will be presented in an expanded companion article \cite{ForresterEdge}.
\section{Notation and the Main Results}\la{sec2}
 \la{sec:Results} The Laughlin state of \(N\) particles  on a cylinder of circumference \( L_x\) at the filling fraction
\(\nu = 1/\beta\) in the Landau gauge is \cite{RH94}
\begin{align}\la{1}
\Psi_{\beta} (z_1,\dots,z_N)= Z_{N}^{-1/2}\Delta^{\beta}
\exp\, ( -\pi\beta\bar\rho \sum_{i = 1}^{N}  y_{i}^{2} ),
\end{align}
where \(z_{i} = x_{i} + i  y_{i}\) is the coordinate on the cylinder, \(\bar\rho=(2\pi\beta\ell_B^2)^{-1}\) is the mean density, \(\ell_{B} = \sqrt{\hbar c/eB}\), \(\) is the magnetic length,   \(Z_{N}\) is the normalization factor and \(\Delta=\prod_{i<j} \left( e^{\ii 2\pi z_{i}/L_x
} - e^{ \ii 2\pi z_{j}/L_x}\right)\). We assume that \(\ell = \sqrt{\beta} \ell_{B}\) is held constant, while varying  \(\beta\), so that the mean density   \(\bar{\rho}  = (2\pi \ell^{2})^{-1}\)  is independent of \(\beta\).
Below, we set the units of length along the cylinder  \(\ell=1\). In these units   the mean density is \( \bar\rho=(2\pi)^{-1}\).

We are interested in the  one-particle density for a  large number of particles
\begin{align}
\label{rhobeta}
\rho_\beta(y)& = N \int\; |\Psi_{\beta}(z,z_2,\dots,z_N)|^{2} \prod^N_{i = 2}dx_idy_i .
\end{align}
 The square of the amplitude of the Laughlin wave function \(|\Psi_\beta|^2\)
can be seen as the Boltzmann weight of a 2D Coulomb plasma
on a cylinder, with neutralizing uniform background charge in the rectangle
$\mathcal R := \{ 0 \le x \le L_x, \:  -\tau L_x \le y \le 0  \}$, where
  \(\tau= N/(\bar\rho L_x^2)\) is an aspect ratio,
and  \(\beta\)   acting as the inverse temperature. The plasma forms a droplet in $\mathcal R$ with approximately uniform density \(\bar\rho\). The mean position of the center of the droplet is given by the exact sum rule\begin{align}
\bar y=N^{-1}\int y\rho_\beta(y)dy= - \frac{L_x \tau}{2}\left(1 - \frac{1}{N}\right).
\end{align}    Outside the droplet, the density decays as a Gaussian. We focus on the density near the right boundary at \(y = 0\), sending the left boundary to negative infinity. This is accomplished by fixing \(\bar{\rho}\), $\tau$, $y$ and sending \(N \rightarrow \infty\).  In this limit, the  shape of the  density is a universal (\(N\), \(\tau\), and \(L_{x}\) independent) function  of $y$. It depends continuously on  \(\beta\) as a parameter. 
The  particle density for different \(\beta\), obtained via simulation, is plotted in Fig. \ref{Fig1}.
The density
approaches \(\bar\rho\) in the bulk.

For \(\beta = 1\), the limiting shape of the density can be computed analytically
from the exact finite $N$ form \cite{CFS83} (see also the sentence below (\ref{Bx})), and we find

\begin{align}
\rho_1(y) = \frac{\bar \rho}{2} \erfc(y) \approx \bar \rho \left\{ 
  \begin{array}{l l}
  1 - (2 \sqrt{\pi} |y|)^{-1} e^{- y^{2}} ,&\,-y \gg 1 \\
\\
  (2 \sqrt{\pi} |y|)^{-1} e^{- y^{2}} ,&  \, y\gg 1 .
  \end{array} \right.\label{rho1}
  \end{align}

We would like to emphasize the Stokes phenomena already seen in this simple case: the density is an entire function having analytically discontinuous asymptotes at different parts of the complex plane \(z\).  

Below (see Eq. \eq{27}) we compute \(\rho_\beta(y)  \) to the leading order in \(\beta-1\), obtaining
the exact functional form of
\begin{equation}
f(y)=\left[ \frac{\partial \rho_\beta}{\bar\rho\partial \beta} \right]_{\beta = 1}
\label{fy}\end{equation} 
The function \(f(y)\) is an entire function. It shows a richer Stokes phenomena: at large \(|y|\)   inside and outside the droplet the density behaves as \begin{align}
 f(y) = \frac{1}{2\sqrt{\pi}}\left\{ 
  \begin{array}{l l}
  3^{5/2}\; y^{-3}e^{-
y^{2}/3},&  \, -y \gg 1 \; (\text{inside})\\
\\
   -|y|\; e^{- y^{2}},&  \, y\gg 1 \; (\text{outside)}.
  \end{array} \right.\label{3}\end{align}
  In fact, the leading  large \(y\)  asymptote outside the  droplet is  independently known for all \(\beta\) \cite{WZ-unpublished,ForresterEdge} up to the coefficient \(c_\beta\). It is
\begin{align}
\label{exterior}
\rho_\beta(y)\approx\bar\rho\,\frac{c_{\beta}}{ \sqrt{2\pi}}(|y|\sqrt{2})^{-\beta} e^{-\beta y^{2}}.
\end{align}
The leading order of the  \(\beta-1\) expansion of the coefficient \(c_\beta\) is found to be \(c_\beta\approx 1  +(\beta-1)(1-{\bf C})/2  + \mathcal{O}(\beta-1)^{2}\), where \({\bf C}\) is the Euler-Mascheroni constant \cite{ForresterEdge}.

The asymptotes of the correction to \(\rho_1(y)\) given by \(f(y)\)  imply an overshoot at
the edge if \(\beta > 1\). This follows by  observing that, when approaching from the inside of the droplet (\(y< 0\)), the correction is positive and rising, while approaching from the outside the correction is negative. A depletion of charge outside the droplet, and accumulation inside is thus implied. Since the total number of particles remains the same, this behavior yields the overshoot. Also, the asymmetry of the asymptotes suggests that the accumulation inside is larger than the depletion outside, indicating a shift of the boundary (where \(f(y)\) vanishes) towards the interior. 

\begin{figure}[h]
\includegraphics[scale=0.45]{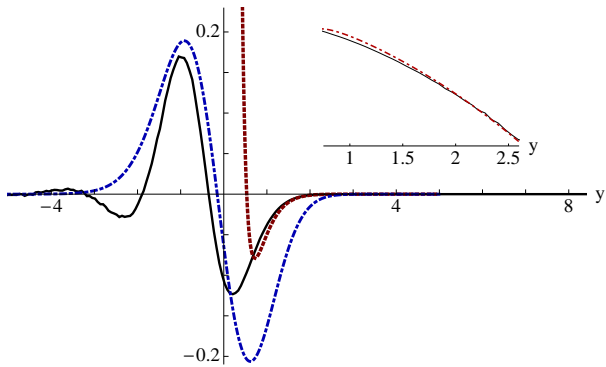}
\caption{Comparison of the numerically computed \( (\rho_\beta - \rho_1)/(\bar\rho(\beta-1))\)
(solid line)
for the Laughlin state \(\nu = \beta^{-1} = 1/3\), with the \(\mathcal{O}(\beta-1)\)
correction \(f(y)\) (blue dash-dotted line), and its exterior asymptote
 Eq. \eq{exterior} (red dashed line)
. The overshoot at the inner edge is captured
well by \(f(y)\). Inset: Log-scale comparison of exterior asymptote to numerical data for the density. Distance is measured in units of \( \ell\).
}
\label{Fig2}
\end{figure}

A noticeable feature of the correction is the  asymmetry of  inside and outside behavior not seen in the integer case where \(\rho_1(-y)=\bar\rho-\rho_1(y)\). Moreover, even deep inside the droplet the \(\beta-1\) correction \({\cal O}(\exp(-y^{2}/3))\) dominates the \(\beta=1 \) contribution \({\cal O}(\exp(-y^{2}))\). This signals that higher order \(\beta-1\) corrections are even more relevant deep inside the droplet. They are responsible  for oscillatory features evident in numerics. 

The full function \(f(y)\) is plotted in Fig \ref{Fig2}, where it is compared with both numerical data for the normalized difference \((\rho_\beta(y) - \rho_1(y))/(\bar\rho(\beta-1))\) at the filling fraction \(\nu = 1/3 \), as well as the exterior asymptote of this difference which follows from equation (\ref{exterior}) with \(\beta = 3\) and the numerical approximation for \(c_{\beta}\) computed in Refs. \cite{Tellez} and \cite{ForresterEdge}. We observe the function \(f(y)\) captures well the overshoot at the inner boundary, as well as the gross overall features of the edge double layer. However, it misses the oscillatory features which arise inside the droplet, since as remarked in the previous paragraph, these are likely non-perturbative corrections in \(\beta - 1\). The exterior asymptote also shows good agreement with the numerics at distances from the edge of order \(\ell\). Therefore, Eqs.(\ref{3}) and (\ref{exterior}) effectively capture the overshoot and approximately illustrate the edge double layer.

Thus, we see that deviations of \(\rho_\beta\) from \(\rho_1\) and subsequently from the step function  are localized to distances of a few magnetic lengths near the edge. In considering the limit of vanishing magnetic length, while keeping the area of the droplet \(2\pi \beta \ell_{B}^{2} N\) fixed, we see the function \(f(y)\) collapses to the \emph{double layer} conjectured in \cite{PW} (see also \cite{Zabrodin, Forrester}). In original  units
\begin{align}\la{7}
\rho_\beta(y) - \rho_1(y) =  \frac{(\beta-1)}{4\pi \beta}  \delta'(y).
\end{align}
This formula succinctly describes the dipole moment per unit length along the edge 

\begin{equation}
\int y\left( \rho_\beta(y) - \rho_1(y)\right) dy = -\frac{\beta-1}{4\pi \beta}
\end{equation}
known independently \cite{Forrester} (see Sec.\ref{sec7} below). 
The dipole moment is completely captured by the $\beta-1$ expansion through the correction $f(y)$. 

Higher moments 
 \( M_{n} \equiv \int_{-\infty}^{\infty}
y^{n} f(y) dy\) grow as \( n!\). 
The odd moments can be obtained explicitly

\begin{align}\la{M}
&M_{2n+1} = - \frac{(2n+1)!}{4^{n}(n+1)!} \left( \frac{1}{2} +  \sum_{k =
1}^{n} R_{k}\right),\\
&R_{k}=\frac{ 3^{k+1}\left(4k+5\right)
+  \left(2k^{2} + 4k + 1\right)
}{16(k+2)(k+1)} . \nonumber 
\end{align}

Finally, in Sec. \ref{sec8} we compute the edge dipole moment per unit length for the Pfaffian states. The result further confirms the close connection between the dipole moment and the Lorentz shear stress alluded to in the introduction. 

\section{Edge Density at $\beta = 1$ }\la{sec3} 
If \(\beta = 1\) \cite{CFS83}, the wave function  (\ref{1}) describes free fermions on the lowest Landau level.   It is the Slater determinant \(\Psi_1={\rm det}  [\psi_{n-1}(x_i,y_i)]_{1\leq n,i\leq N} \) of   one-particle wave functions \(\psi_{n}(z)=\left(L_x \sqrt{\pi}\right)^{{-1/2}}
e^{-\frac{1}{2}(y+ p_n)^2}e^{ip_nx}\), where \(p_n=2\pi n L_x^{-1},\;n=0,\dots,N-1\). Consequently,  the  \(n\)-point function\begin{align}g_1^{(n)} \equiv  \frac{N!}{(N-n)!}\int |\Psi_{1}(z_{1},...,z_{N})|^2\prod_{i
= n+1}^{N}d^2z_{i}\la{g} \end{align} is also a determinant (per Wick's theorem) \( g_1^{(n)} =\det \left[  K(z_i,z_j)\right]_{i,j = 1,...,n}
\). Here \be\la{K} K(z_{1},{z_2}) =  \sum_{n = 0}^{N-1} \psi_{p_n}(z_1) \psi_{p_n}^{*}(z_2) \ee
 is the Fredholm kernel (projector) over the Landau level.  We will use
the Fourier modes of the kernel along the boundary
\begin{align}&K(z_1,z_2)=L_x^{-1}\sum_{n=0}^{N-1} K_{p_n}(y_1,y_2)e^{ip_n(x_1-x_2)},\\
&K_p(y_1,y_2)=\frac{1}{\sqrt{\pi}}e^{-\frac
1 2 (y_1+ p)^2}e^{-\frac 1 2 (y_2+ p )^2}. \label{Bx}
\end{align}
In particular, for the density we have \(\rho_1(y)=L_x^{-1}\sum_pK_p(y,y)\). The limiting shape of the density is given by  (\ref{rho1}).
\smallskip
\section{Perturbation expansion in $\beta - 1$}\la{sec4}
%
As \(\beta\) departs from 1, the wave function and correlation functions  depart from the determinantal form. This can be viewed as an interaction between fermions.   We can develop a perturbation expansion of the density around \(\beta = 1\) by direct expansion of (\ref{rhobeta}) in \(\beta-1\). 

Defining the expectation value over free fermions \( \langle \mathcal{A}\rangle =\int\Psi_{1}^{\ast}\mathcal{A}\Psi_1 \prod_{i = 1}^{N}dx_{i}dy_{i}\), where \(\mathcal{A} \) is a symmetric function of coordinates, we write the first correction to the  unperturbed density  \(\rho_1(y)\) as
\begin{align}\la{f}
f(y)=-2\pi  \left\langle  \sum_i\delta^{(2)}(z - z_{i})[   \sum_{ j>k} V(z_j,z_k)+\sum_j y_j^2] \right\rangle_{c}
\end{align}
where
\begin{align}
V(z,z')=-\log \left|e^{ 2\pi\ii z /L_x} - e^{2\pi\ii z'/L_x}\right|^{2}.
\end{align}
The symbol \( \langle ... \rangle_{c} \) stands for the connected part of the correlation function defined for two operators as \(\langle A B\rangle_{c} =\langle AB\rangle - \langle A\rangle\langle B\rangle\). In this case, it involves 1-, 2- , and 3- point correlation functions, which can be computed using Wick's theorem. The result is expressed through the 
two-body interaction potential \(\varphi_1(z)= \int d^{2}z' \rho_1(z')V(z,z')+y^2\) and the two-body potential \(V(z,z') \). We can write both contributions together by introducing a non-symmetric potential
 \begin{align}
v(z_1,z_2)=-\varphi_1(z_1)+ V(z_1,z_2).\quad 
\end{align} 
The result  reads
\begin{align}
f(y)=2\pi \int d^{2}z' d^{2}z'' K(z,z') K(z',z'')K(z'',z)& \times \nonumber\\ [
v(z,z')-v(z',z'')  ].&\end{align}

We comment that perturbation theory around the integer filling factor  could be carried out from the phenomenological Hamiltonian of the FQHE  considered in  \cite{PW}.   We briefly describe it. 

The Hilbert space of FQH states is comprised of functions \(Q\Psi_1 \), where \(Q=Q_k(e^{2\pi\ii  z_1/L_x},\dots , e^{2\pi\ii  z_N/L_x}) e^{-\frac k2 \sum_{i=1}^N y_i^2}\),  and \(Q_k\) is a symmetric polynomial of the degree \(k-1\).
  In particular, Laughlin's wave function corresponds to $k=\beta-1$ and $Q_k=\Delta^{\beta-1}\). The operators acting in this space are symmetric functions of coordinates and derivatives. Their  matrix elements   are $\langle Q|{\cal A}|Q'\rangle=\int |\Psi_1|^2 Q{\cal A}Q' \prod_i d x_idy_i$, where derivatives in \({\cal A}   \) (if any) act to the right. Magnetic translation operators in this representation are  $$\pp_i=\p_{z_i}-(\beta-1)\p_{z_i}[y_i^2+\sum_{j\neq i}V(z_i,z_j)].$$ Each operator annihilates the Laughlin state \(\pp_i\Psi_\beta=0\). 

The phenomenological Hamiltonian  is constructed as a square of magnetic translations $H=\sum_i \pp_i^\dag \pp_i$. Its spectrum is non-negative, with the Laughlin wave function as the zero energy ground state. The interaction is explicit in the Hamiltonian. The formula \eq{f} could be obtained as  the leading order of the perturbation expansion.


 Further calculations are most conveniently done using Fourier modes. 
 We write the potential as\begin{align}\nonumber
v(z,z')= v_0(y,y')+\sum_{n = 1}^{\infty} 2 \cos\left(k_n(x'-x)\right) v_{n}(y-y'),\end{align}where
\begin{align}\nonumber
&\quad v_n(y)=\frac {2\pi}{L_xk_n}e^{-k_n  |y | },\;v_0(y,y')=\frac{4\pi}{L_x}{\rm
min}(y,y')-\varphi_1(y) .
\end{align} 
Then
\begin{align}
&f(y)= \frac{2\pi}{L_{x}}\sum^{N-1}_{ n,m=0 } \int dy'dy'' K_{p_n}(y,
y') K_{p_m}(y', y'') )\times\nonumber\\
&[K_{p_m}(y'',y)v_{|n\!-\!m|}(y,y')\!-\!K_{p_n}(y'',y)v_{|n\!-\!m|}(y' , y'') ].
\end{align}
The integrals in \(y\) can be computed analytically. The result is expressed through the error function \({\rm erf}(y)=\frac{2}{\sqrt\pi}\int_0^y e^{-y'^2}dy',\;{\rm erfc}(y)=1-{\rm erf}(y)\) and its antiderivative \(F(y)
= -y\, {\rm erfc}(y) + e^{- y^{2}}/\sqrt{\pi},\;F'(y)=-{\rm erfc}(y)\). 

We list the contributions of the zero and non-zero modes separately \(f(y)=f^{(1)}(y)+f^{(2)}(y)  \)
\begin{align}\la{zero}
f^{(1)}(y) =   &2\pi (y\rho_1)'+
\frac{\pi}{2} \rho_1''(y)+
 \frac{1}{\sqrt{\pi}}\left(\frac{2\pi}{L_x}\right)^{2}\times\\ \sum_{0 \le n\ne m< N}&e^{- \left(y +  p_n\right)^{2}} \left [F\left(y+p_m\right) - \sqrt{2} F\left(\frac{p_m - p_n}{\sqrt{2}}\right)\right],\nonumber\end{align}\begin{align}
f^{(2)}(y)  = \frac{1}{\sqrt{\pi}}\left(\frac{2\pi}{L_x}\right)^{2}& \sum_{0 \le n\ne m< N}\frac{e^{-(y
+ p_n)^{2}}}{p_n-p_m}\times\la{nonzero}\\& \left[{\rm erf}\left( y+ p_m\right) + {\rm erf}\left(\frac{
p_n - p_m}{\sqrt{2}}\right) \right]\nonumber .
\end{align}
This formula could be studied in various different large \(N\) limits, while \(y \) is of the order 1. One is the limit of the thin cylinder with the large aspect ratio \(\tau\gg 1\), when the circumference \(L_x\) is comparable with the magnetic length. 
In this case electrons are crystallized all the way into the bulk \cite{SWK04}\cite{CFS83}\cite{JLS09}.
 The density is a nearly periodic function featuring \(N\) humps with a width of the magnetic length localized at \(y_n=2\pi n/L_x\).
The contribution comes only from the zero mode \eq{zero}.
In
this case, the correction we computed does not drastically change the profile away from \(\beta=1\).

However for a cylinder with
aspect ratio of order 1, the genuine 2D case, the distance between humps and the size of the humps are of the same order. The sums in (\ref{zero},\ref{nonzero}) must be replaced by integrals. This gives the limiting shape
\begin{align}\la{27}
f^{(1)}(y) = \frac 12 \erfc (y)&-\frac{y}{2\sqrt\pi}e^{-y^2}+
 \\
 \frac{1}{\sqrt{\pi}}\iint_{y}^\infty e^{-p^{2}} &\left [F (p') - \sqrt{2}
F\left(\frac{p' - p}{\sqrt{2}}\right)\right]dpdp'\nonumber\\
f^{(2)}(y)  =  \frac{1}{\sqrt\pi} \iint_{y}^\infty &\frac{e^{-p^{2}}}{p-p'}\left[{\rm
erf}\,   p' + {\rm erf}\left(\frac{
p-p'}{\sqrt 2}\right)\right]dpdp'  . \nonumber
\end{align}
The formulas are more transparent if we focus on the antisymmetric part of
the correction \(f_A(y)=\frac{1}{2}(f(y)-f(-y))\).
The  result is more compact if one applies the differential
operator \(D=(\p_y+2y)\p_y\) to the antisymmetric part.  This
operator annihilates the zero order part of the density \(D\rho_1=0\). After some tedious algebra we get
\begin{align}
\!Df_A\!=\!\left(y^{2}\!-\!\frac 12\right) \left({\rm erf}\left(\frac{y}{\sqrt{3}}\right)\!-\!{\rm erf}( \!\,
y\,\! )
\right)  + \frac{\sqrt{3}}{\sqrt{\pi}} y e^{-\frac{y^{2}}{3}}.\la{30}
\end{align}
The odd moments \eq{M} were computed from this formula.

\section{Asymptotes of the Overshoot}\la{sec5}
These formulas are sufficient to compute the asymptotes of the overshoot at large \(|y|\) (larger than magnetic length but smaller
than the circumference of the boundary). They are  summarized in Sec.\ref{sec2}.

The asymptotes outside of the droplet are readily computable directly from
the integrals (\ref{27}). In this case, the integrals in \eq{27} are suppressed by the factor \(e^{-p^2}\). We reproduce the \(\beta-1\) expansion of the known asymptote \eq{exterior} and obtain the \(\beta-1\) expansion of the previously unknown coefficient \(c_\beta\) \cite{ForresterEdge}. 

Inside the droplet \eq{30} indicates that the dominant term comes from  \({\rm erf}(y/\sqrt{3})\) and the last term  in \eq{30}. It is \({\cal\ O}(e^{-y^2/3})\) and therefore dominates other contributions of the order \({\cal\ O}(e^{-y^2})\). As a result, inside  the droplet    \(f\approx  2 f_A\). The asymptotes of \(f_A\) easily follow from \eq{30}. The result is presented in \eq{3}.

We also want to comment further on the emergence of the curious \( e^{-y^{2}/3}\)
asymptote only inside the droplet. A similar effect is seen in the \(\beta
- 1\) correction to the pair distribution function \(g_\beta(r_1,r_2)\)  deep inside the bulk where it depends only on the distance between the points.   At \(\beta=1\) the pair correlation function is \(g_1(r) = 1- e^{- r^{2}/2}\) as it follows from (\ref{g},\ref{K}). However, the correction computed in \cite{Jancovici} has the asymptote \(\sim r^{-2}\exp\left(- r^{2}/4\right)\). In that case, the appearance of the
different exponent can be understood as necessarily following from the functional rule \(g_\beta( r) =-  e^{- r^{2}/2} g_\beta(\ii r)\)  for  the pair correlation function \cite{Samaj}. To the best of
our knowledge, there is no analogous sum rule to explain the \(\exp\left(-y^{2}/3\right)\)
decay for the density. We expect that higher order corrections show even slower decay, indicating the development of oscillations extended into the bulk.

The apparent asymmetry between asymptotes inside and outside 
the droplet and a decay \({\cal\ O}(e^{-y^2/3})\) are perhaps the major results
of the paper.

\section{Singular Double Layer}\la{sec6}
To capture the singular character of the edge, we consider the different limit when \(\ell\to 0\). In this case, the limiting shape collapses to  a singularity at the edge. Let us restore the magnetic length in the formulas \eq{M}, fix \(y\) and send magnetic length to zero. In  the original coordinates the moments \(\int (\rho_\beta(y)-\rho_1(y))y^ndy=(2\pi)^{-1}(\beta-1)\ell^{n-1}M_n\) are scaled as  \( \ell^{n-1} \).  In the limit \(\ell\to 0\), the only moment which remains is the dimensionless dipole moment \( M_{1} =-1/2\). It is saturated  by the double layer (\ref{7}).

We want to emphasize that this is an fundamental feature of the Laughlin state that persists in the thermodynamic (large $N$) limit. In taking the the limit of $N \to \infty$ and $\ell \to 0$, the droplet tends to a uniform density on a domain with a step-function support. However, the magnitude of the dipole moment per unit length along the boundary is a universal function of the filling fraction, and survives in the large $N$ limit. Since the dipole moment is localized to distances $\mathcal{O}(\ell)$ near the edge, this gives rise to a singular double layer correction at the boundary of the droplet which is {\it essential} to the physics of the FQH state.

 \section{Dipole moment from Exact Sum Rule in Disk geometry}\la{sec7}

Though we have specialized the preceding discussion to the case of a cylindrical geometry, the edge density derived above also follows from studying the Laughlin state in the disk geometry in local coordinates, and under the proper scaling limit, in which the droplet boundary appears approximately flat (vanishing geodesic curvature). This is discussed in great detail in the companion paper \cite{ForresterEdge}. This universality of the edge density can be understood as arising from the finite correlation length in the bulk, wherein correlations show Gaussian decay over a characteristic distance on the order of the magnetic length. Thus, the shape of the droplet is invisible to the edge density, which is defined for distances from the boundary on the order or larger than $\ell$, but much less than the droplet size.


In this section, we present an alternative derivation of the edge dipole moment from an exact sum rule for the droplet in the disk geometry. This was obtained in \cite{Forrester}, and we repeat the arguments below. In the next section, we use this approach to compute the dipole moment of Pfaffian FQH ground state wave functions. 

To begin, we write the Laughlin wave function on the plane in the symmetric gauge with the same scaling of coordinates as above


\begin{equation}
\Psi_{\beta}(z_{1},..., z_{N}) = Z_{N}^{-1/2} \prod_{i<j} (z_{i} - z_{j})^{\beta} e^{- \frac{1}{2}\pi \beta \bar{\rho} \sum_{i} |z_{i}|^{2}}. 
\end{equation}
The droplet now achieves approximately uniform density $\bar{\rho}$ over a region on the plane bounded by the disk of radius $R = \sqrt{N/\pi \bar{\rho}}$. We can compute exactly the second moment of the density, using the following arguments. Under a rescaling of coordinates $z_{i}\to \lambda z_{i}$, the normalization integral is unchanged, which implies $\partial_{\lambda} \log Z_{N}(\lambda) = 0$. The derivative with respect to $\lambda$ can then be carried out explicitly. The scaling of the Laughlin-Jastrow factor, as well as the integration measure, produces a factor $\beta N(N-1) + 2N$, while differentiation of the Gaussian exponential factor will result in the expectation value of $\sum_{i} |z_{i}|^{2}$. Then, setting $\lambda = 1$, we recover the exact sum rule

\begin{equation}\la{sumrule1}
\int |z|^{2}  \rho_{\beta}(z) d^{2} z = \frac{N}{2\pi \bar{\rho}}\left( N +  \frac{2 - \beta}{\beta}\right).
\end{equation}
Subtracting off the density at $\beta = 1$, this becomes

\begin{equation} \la{sumrule2}
\int |z|^{2}  \left(\rho_{\beta}(z)  - \rho_{1}(z)\right) d^{2} z = \frac{N}{\pi \bar{\rho}}\left(  \frac{1 - \beta}{\beta} \right).
\end{equation}
We are interested in the contribution to this sum rule which comes from the edge double layer. Consider a change of coordinates $r = R + y$, and take $R, N \to \infty$ while keeping $\bar{\rho}$ fixed. In this limit, we recover the edge density above as a function of $y$, the distance from the boundary. Now, we expand the LHS of (\ref{sumrule2}) in $R$, and use

\begin{align}
2\pi \int_{-R}^{\infty} (R + y) \left(\rho_{\beta}(y)  - \rho_{1}(y)\right) dy = 0,
\end{align}
which is just the statement that both densities $\rho_{\beta}$ and $\rho_{1}$ count the same total number of particles. Dividing through by $N$ and sending $R, N \to \infty$ then gives 


\begin{align}
 \int_{-\infty}^{\infty}   y \left(\rho_{\beta}(y)  - \rho_{1}(y)\right) dy  = - \frac{1}{4\pi }\left(  \frac{\beta - 1}{\beta} \right).
\end{align}
Thus, the dipole moment is recovered, and shown to be linear in $\nu - 1$. This explains the success of the perturbative expansion at order $\beta - 1$ in capturing the double layer. 

\section{Edge Dipole Moment of Pfaffian states}\la{sec8}
We can derive the same result for the Pfaffian states, defined by the ground state wave function

\begin{align}\la{Pf}
\Psi_{\beta}^{pf} = {\rm Pf} \left( \frac{1}{z_{i} - z_{j}}\right) \Delta^{\beta} e^{- \frac{1}{2}\pi \beta \bar{\rho} \sum_{i} |z_{i}|^{2}}  .
\end{align}
Here ${\rm Pf}(M_{ij})$ is the Pfaffian of the matrix $M$ \cite{Moore91}. Under this scaling, the Pfaffian states describing filling fraction $\nu = 1/\beta$ will have a mean density $\bar{\rho}$ in the bulk. For $\beta = 2$, this is the well-known Moore-Read state believed to describe the FQH effect at filling fraction $\nu = 5/2$. Following the arguments in the preceding section, we find the second moment of the density for the Pfaffian state 

\begin{equation}\la{sumrulepf}
\int |z|^{2} \rho_{\beta}^{pf} (z) d^{2} z = \frac{N}{2\pi \bar{\rho}} \left( N + \frac{2 - \mathcal{S} }{\beta}\right).
\end{equation}

The quantity $\mathcal{S} = \beta + 1$ for Pfaffian states, and is known as the ``shift" for FQH states. The shift relates the number of electrons in a FQH state needed to cover a sphere pierced by $N_{\phi}$ flux quanta, and is defined by the equation $N_{\phi} = \nu^{-1} N - \mathcal{S}$ \cite{shift}. Normalizability of the many-particle wave function on the sphere requires the holomorphic part of the wave function to be homogeneous of total degree $N N_{\phi}/2$. Comparing this with the total degree of the Pfaffian wave function $ N(\beta N-\beta - 1)/2 $ determines the shift \cite{ReadRezayi11}, and accounts for its appearance in the second moment sum rule \eq{sumrulepf}. Importantly, the shift determines the Lorentz shear modulus (odd or Hall viscosity) $\eta^{(A)} = \hbar \bar{\rho} \mathcal{S}/4$ \cite{Read09}. Following the steps above, we deduce that the dipole moment for the Pfaffian state is, upon subtracting the $\beta = 1$ (free fermion) density for reference,

\begin{align}\la{dipolemompf}
\int y \left(\rho_{\beta}^{pf} (y)  - \rho_{1}(y)\right) dy = - \frac{1}{4\pi} \left( \frac{\mathcal{S} - 2}{2\beta} + \frac{1}{2}\right). 
\end{align}
Note that for Pfaffian states \eq{Pf} with $\beta \ge 1$, the dipole moment is negative, indicating the presence of an overshoot at the edge. Furthermore, the dependence on $\mathcal{S}$, and not simply the filling fraction, confirms a connection between the strength of the dipole moment and the odd viscosity \cite{FDMH09}. This connection is somewhat obscured in the Laughlin state for which $\mathcal{S} = \beta$. 

While the result \eq{dipolemompf} is specific to Pfaffian states at filling fraction $\nu = \beta^{-1}$, it can be readily adapted to other FQH model wave functions. Indeed, the derivation requires that the ground state wave function in the symmetric gauge consists of a product of Gaussian exponential factors multiplied by a homogeneous function of particle coordinates of total degree $N N_{\phi}/2 = N (\nu^{-1} N - \mathcal{S})/2$. These are rather generic features of FQH model wave functions in the lowest Landau level, and suggest the dependence of the edge dipole moment \eq{dipolemompf} on the shift $\mathcal{S}$ is a general property of FQH states. 

\section{Conclusion}
In this article, we studied the functional form of the edge density of the Laughlin state at filling fraction \(\nu = \beta^{-1}$ using a perturbative expansion around the free fermion, IQH state at $\beta = 1$. From this we obtained the exact, finite $N$ correction to the density at order $\beta - 1$. The correction becomes a universal function in the limit of large $N$, and we explicitly compute the odd moments and obtain exact asymptotics of the edge density. Our results confirm the presence of the dipole moment localized on the order of $\ell$ near the edge, which collapses to a singular double layer in the limit of large magnetic field. We have demonstrated that the perturbative result effectively captures the double layer structure of the edge, even for larger values of $\beta$, although it cannot account for the non-perturbative oscillatory features. Finally, we presented a route to analyze the edge structure of different FQH states, and derived the edge dipole moment of the Pfaffian state. The results reveal a relationship between the edge dipole moment and the bulk odd viscosity.

P.W. and T.C. thank A. Abanov, A. Zabrodin and I. Gruzberg  for numerous discussions. The work of P.W. and T.C.  was supported by NSF DMS-1206648, DMR-0820054 and BSF-2010345.
P.W. thanks Andrea Cappelli for the hospitality at  INFN Florence when this paper was completed. G.T. acknowledges financial support from Fa\-cultad de Ciencias, Uniandes.


\begin{thebibliography}{99}
\bibitem{Shytov} We used numerical data obtained through Monte Carlo techniques by A. Shytov for $N = 200$ particles.
\end{thebibliography}
\end{document}